\documentclass{article}

\usepackage{epsfig}
\usepackage{subfigure}
\usepackage{calc}
\usepackage{graphicx}

\usepackage{amsfonts}
\usepackage{amsmath}
\usepackage{amssymb}
\usepackage{latexsym}

\newtheorem{theorem}{Theorem}[section]

\newtheorem{example}[theorem]{Example}

\newtheorem{definition}[theorem]{Definition}

\newtheorem{proposition}[theorem]{Proposition}

\begin{document}

\title{A Simple Computational Model for Acceptance/Rejection of Binary Sequence Generators}
\date{}
\author{Amparo F\'{u}ster-Sabater$^{(1)}$ and Pedro Garc\'{\i}a-Mochales$^{(2)}$\\
{\small (1) Instituto de F\'{\i}sica Aplicada, C.S.I.C.}\\
{\small Serrano 144, 28006 Madrid, Spain} \\
{\small amparo@iec.csic.es}\\
{\small (2) Instituto de Ciencia de Materiales de Madrid, C.S.I.C.} \\
{\small Cantoblanco. 28049 Madrid, Spain} \\
{\small pedro.garciamochales@uam.es }}

\maketitle

\begin{abstract}

A simple binary model to compute the degree of balancedness in the
output sequence of LFSR-combinational generators has been
developed. The computational method is based exclusively on the
handling of binary strings by means of logic operations. The
proposed model can serve as a deterministic alternative to
existing probabilistic methods for checking balancedness in binary
sequence generators. The procedure here described can be devised
as a first selective criterium for acceptance/rejection of this
type of generators.

Keywords: Balancedness, Bit-string model, Combinational generator,
Design rules

\end{abstract}

\section{Introduction}
\footnotetext{Work supported by Ministerio de Educaci\'{o}n y
Ciencia (Spain) Projects SEG2004-02418 and SEG2004-04352-C04-03.\\
Applied Mathematical Modelling. Volume 31, Issue 8, pp. 1548-1558. August 2007. \\
DOI:10.1016/j.apm.2006.05.004 } \noindent Pseudorandom binary
sequences are simple tools with application in fields so different
as spread-spectrum communications, circuit testing,
error-correcting codes, numerical simulations or cryptography.
Most generators producing these sequences are based on
combinational Boolean functions \cite{Man} and Linear Feedback
Shift Registers (LFSRs) \cite{Go}. These combinational generators
(in the sequel LFSR-combinational generators) can include an
unique LFSR (\textit{nonlinear filters}) or more than one LFSR
(\textit{nonlinear combination generators}). In both cases, the
\textit{output sequence} is a periodic sequence generated as the
image of a nonlinear Boolean function $F$ in the LFSR cells (see
\cite{Ru}, \cite{Si}).

Balancedness in the generated sequence is one necessary condition
that every LFSR-combinational generator must satisfy. Roughly
speaking, a binary sequence is balanced whether it has
approximately the same number of $1^{\prime }s$ as $0^{\prime }s$.
Due to the long period of the sequences produced by
LFSR-combinational generators (for instance, $T \simeq 10^{38}$
bits in current cryptographic applications), it is unfeasible to generate
an entire cycle and then count the number of $1^{\prime }s$ and $%
0^{\prime }s$. Therefore, in practical design of binary
generators, portions of the output sequence are chosen randomly
and statistical tests (e.g. frequency test or monobit test
\cite{Me}) are applied to all these subsequences. Nevertheless,
passing the previous tests merely provides \textit{probabilistic
evidence} that the generator produces a balanced sequence.

In the present work, balancedness of LFSR-combinational generators
has been treated in a \emph{deterministic way}. In fact, a simple
binary model allows one to compute the exact number of $1^{\prime
}s$ in the output sequence without producing the whole sequence.
From such a model, the general expression of the number of
$1^{\prime }s$ is obtained as a function of the generator
parameters. In this way, the degree of balancedness of such a
sequence can be perfectly checked: the obtained number of
$1^{\prime }s$ is compared with the value required for this
sequence to be balanced (half the period $\pm$ a tolerance
interval). In case of non-accordance, the LFSR-combinational
generator must be rejected. Thus, the procedure here developed can
be considered as a first selective criterium for
acceptance/rejection of this type of generators.

The computational method is based exclusively on the handling of
binary strings by means of logic operations (OR, AND). Indeed, the
general expression of the number of $1^{\prime }s$ is just an
interpretation of such binary strings. As a straight consequence
of this method, practical rules to design generators with balanced
(quasi-balanced) output sequences are also derived. The procedure
can be applied to LFSR-combinational generators in a range of
practical interest. Some illustrative examples including filters
and combination generators complete the work.

\section{Fundamentals and Basic Concepts}
Several basic concepts and definitions to be used throughout the
paper will be presented in the following subsections. First of
all, the concept of \textit{minterm} is introduced:
\begin{definition}
A minterm of L binary variables $(m_{0},\,m_{1},\,...,\,%
\,m_{L-1})$ is a monomial of the $L$ variables, where each
variable can be in its true or complementary form. For L
variables, there exist $2^{L}$ minterms, each minterm being
expressed as the logic product of the L (appropriately
complemented) variables.
\end{definition}
A minterm of $L$ variables is denoted by $M_{\alpha}$ where
$\alpha$ is a binary configuration of $L$ bits. The \textit{i-th}
bit (numbered from right to left) is $1$ if the variable $m_i$ is
in true form and $0$ if the variable $m_i$ is in complementary
form. Since a binary sequence generator is characterized by a
Boolean function, the representation of such functions is
considered.
\subsection{Representation of Boolean Functions}
Two different representations of Boolean functions are introduced.
\begin{enumerate}
    \item \textit{Boolean functions in Algebraic Normal Form:} Any
$L$-variable Boolean function can be uniquely expressed in
Algebraic Normal Form (ANF) or M\"{u}ller expansion (\cite{Man},
\cite{Wang}) by means of the exclusive-OR sum of logic products in
the $L$ variables. A simple example of Boolean function in ANF is:
\[
F(m_{0},\,m_{1},\,...,\,%
\,m_{L-1})=m_{L-1}\,m_{1}\,m_{0}\oplus m_{L-2}\,m_{1}\oplus
m_{L-1},
\]
where the concatenation of variables represents the logic product
and the symbol $\oplus $ the exclusive-OR logic operation.
    \item \textit{Boolean functions in terms of their minterms:} Any
$L$-variable Boolean function can be canonically expressed as a
linear combination of its minterms (\cite{Man}, \cite{Wang}). A
simple example of Boolean function in terms of its minterms is:
\[
F'(m_{0},\,m_{1},\,...,\,%
\,m_{L-1})=M_{10 \ldots 01}\oplus M_{01 \ldots 10}\,.
\]
In the term $M_{10 \ldots 01}=m_{L-1}\, \overline
{m}_{L-2}\,\ldots \, \overline {m}_{1} \, m_{0}$ only the
variables $m_{0}$ and $m_{L-1}$ are in true form while in the term
$M_{01 \ldots 10}= \overline m_{L-1}\, {m}_{L-2}\,\ldots \,
{m}_{1} \, \overline m_{0}$ only the variables $m_{0}$ and
$m_{L-1}$ are in complementary form.
\end{enumerate}

In this work, both representations of Boolean functions will be
systematically addressed.

\subsection{LFSRs and Minterm Functions}
A binary LFSR is an electronic device with $L$ memory cells
(stages), numbered $0,1,...,L-1$, each of one capable of storing
one bit. The binary content of the $L$ stages at each unit of time
is the \textit{state} of the LFSR at that instant. In addition, a
clock controls the shift of data. At each unit of time the
following operations \cite{Go} are performed: (i) The content of
stage $0$ is output ; (ii) the content of stage $i$ is moved to
stage $i-1$ for each $i$, $1\leq i\leq L-1$ ; (iii) The new
content of stage $L-1$ is the exclusive-OR of a subset of stages
given by $P(x)$, that is the LFSR connection polynomial. If $P(x)$
is a \textit{primitive polynomial} of degree $L$ \cite{Li}, then
the LFSR is called a maximum-length LFSR. In the sequel, only
maximum-length LFSRs will be considered.

\begin{definition}
An LFSR-combinational generator is a nonlinear Boolean function
$F$ defined as $F:GF(2)^L-\{0\}\rightarrow GF(2)$, whose input
variables $m_{i}\;(i=0,...,L-1)$ are the binary contents of the
LFSR stages.
\end{definition}
At each new clock pulse, the new binary contents of the stages
will be the new input variables of the function $F$. In this way,
the generator produces the successive bits of the output sequence.
An LFSR-combinational generator is a nonlinear Boolean function
$F$ given in its ANF.

\begin{definition}
A minterm function is a minterm of $L$ variables expressed in ANF.
Every minterm function can be easily obtained by expanding out the
corresponding minterm.
\end{definition}

Let $M_{\alpha}$ be a minterm of $L$ variables where $\alpha$
includes $d$ $1^{\prime }s$ in its binary representation with
$\;(1\leq d\leq L)$. According to \cite{Man}, its corresponding
minterm function is perfectly characterized:
\begin{itemize}
    \item Such a minterm function has ${L-d \choose 0}$ terms of order
    $d$, ${L-d \choose 1}$ terms of order $d+1$, ${L-d \choose 2}$
    terms of order $d+2$, $\ldots$, ${L-d \choose L-d}$ terms of order
    $L$, so in total the number of terms is:
    \begin{equation}\label{equation:1}
    \mbox{No. of terms} = 2^{L-d}.
    \end {equation}
    \item The particular
    form of the terms of each orden is perfectly determined too, see
    \cite{Man}.
\end{itemize}

On the other hand, every minterm function considered as an
combinational generator applied to the $L$ stages of an LFSR
generates a canonical sequence with a unique $1$ and period
$T=2^{L}-1$ (see \cite{Ru}). The location of this $1$ depends on
the LFSR initial state. Let us see for a simple example the
particular form of the minterm functions and their corresponding
canonical sequences.

\begin{example}For a LFSR of $L=3$
stages, connection polynomial $P(x)=x^3+x^2+1$ and initial state $%
(1\,1\,0) $\, we have:
\[
\begin{tabular}{ll}
$M_{111}=m_{2}\,m_{1}\,m_{0}$ & $\longleftrightarrow
\;\;\{0,0,0,0,0,0,1\}$
\\
$M_{110}={m}_{2}\,m_{1}\,\overline{m}_{0}=m_{2}\,m_{1}\,m_{0}\oplus
m_{2}\,m_{1}\,$ & $\longleftrightarrow
\;\;\{0,0,0,0,0,1,0\}$ \\
$M_{101}=m_{2}\,\overline{m}_{1}\,m_{0}=m_{2}\,m_{1}\,m_{0}\oplus
m_{2}\,m_{0}\,$ & $\longleftrightarrow
\;\;\{0,0,0,0,1,0,0\}$ \\
$M_{010}=\overline{m}_{2}\,m_{1}\,\overline{m}_{0}=m_{2}\,m_{1}\,m_{0}\oplus
m_{2}\,m_{1}\oplus m_{1}\,m_{0}\oplus m_{1}\,$
& $\longleftrightarrow \;\;\{0,0,0,1,0,0,0\}$ \\
$M_{100}={m}_{2}\,\overline{m}_{1}\,\overline
{m}_{0}=m_{2}\,m_{1}\,m_{0}\oplus m_{2}\,m_{1}\oplus
m_{2}\,m_{0}\oplus m_{2}\,$
& $\longleftrightarrow \;\;\{0,0,1,0,0,0,0\}$ \\
$M_{001}=\overline{m}_{2}\,\overline{m}_{1}\,m_{0}=m_{2}\,m_{1}\,m_{0}\oplus
m_{2}\,m_{0}\oplus m_{1}\,m_{0}\oplus m_{0}\,$
& $\longleftrightarrow \;\;\{0,1,0,0,0,0,0\}$ \\
$M_{011}=\overline{m}_{2}\,m_{1}\,m_{0}=m_{2}\,m_{1}\,m_{0}\oplus
m_{1}\,m_{0}\,$ & $\longleftrightarrow
\;\;\{1,0,0,0,0,0,0\}$.%
\end{tabular}
\]
The left column represents the minterms and their corresponding
minterm functions while the right column shows the generated
sequences when the LFSR cycles through its $2^3-1$ states
\[
(1\,1\,0), (1\,0\,0), (0\,0\,1), (0\,1\,0), (1\,0\,1), (0\,1\,1),
(1\,1\,1).
\]
In this way, an arbitrary combinational generator e.g.
$F=M_{001}\oplus M_{010}\oplus M_{100}$ applied to the $3$ stages
of the previous LFSR will produce the sequence
$\{0,1,1,1,0,0,0\}$. That is to say, the bit-wise exclusive-OR of
the canonical sequences associated to the corresponding minterms.
Remark that the minterm with all the variables in complementary
form is excluded from the succession of minterms as well as the
state all zeros is excluded from the succession of LFSR states
\cite{Go}.
\end{example}

Let us now generalize the concept of minterm function when more
than one LFSR are involved.

\subsection{Generalization of the Minterm Functions to more than one LFSR}
The generalization of the previous concepts to several LFSRs is
quite immediate. In fact, let $A,\,B,...,\,Z$ be maximum-length
LFSRs whose lengths are respectively $L_{A},$
$\,L_{B},\,...\,,\,L_{Z}$ (supposed $(L_i,L_j)=1,\; i\neq j$). We
denote by $a_{i}\;(i=0,...,L_{A}-1),\,$
$b_{j}\;(j=0,...,L_{B}-1),\,...\,, \,z_{k}$ $(k=0,...,L_{Z}-1)$
their corresponding stages. The minterms of a nonlinear
combination generator, called \textit{global minterms}, are of the
form, e.g.
\[A_{ij}\,B_{pqr}\,...\,Z_{s},\]
that is the logic product of the individual minterms of each LFSR.
Therefore, each global minterm depends on $L$ variables, where $L$
is given by:
\[ L=L_{A}+\,L_{B}+\,...+%
\,L_{Z}. \] A global minterm function is a global minterm of $L$
variables expressed in ANF As before every global minterm function
considered as a combinational generator applied to the stages of
the LFSRs \cite{Ru} produces a canonical
sequence with an unique $1$ and period $%
T=(2^{L_{A}}-1)(2^{L_{B}}-1)\,...\,(2^{L_{Z}}-1)$. The location of
this $1$ depends on the initial states of the LFSRs.

In brief, every LFSR-combinational generator can be expressed as a
linear combination of its minterms as well as each minterm
provides the output sequence with a unique $1$. Thus, the basic
idea of this work can be summarized as follows:

The number of minterms in the expression of $F$ equals the number
of $1^{\prime }s$ in the output sequence.

As every LFSR-combinational generator is designed in Algebraic
Normal Form, the Boolean function $F$ has first to be converted
from its ANF into its minterm expansion. Then, the computation of
minterms will give us the corresponding number of $1^{\prime }s$
in the output sequence and consequently its exact degree of
balancedness.

\subsection{Conversion Method from Algebraic Normal Form into
Minterm Expansion} In order to carry out this conversion, the
following function that maps Boolean functions to their dual
functions is defined:
\begin{definition}
Let $F$ be a Boolean function given in ANF. The function $\Phi_F$
is defined as a Boolean function that substitutes each term
$m_{i}\,m_{j}\, ...\,m_{k}$ of $F$ for its corresponding minterm
$M_{\alpha}$ where $\alpha$ includes $1's$ at the \textit{i-th},
\textit{j-th}, $\ldots$, \textit{k-th} positions.
\end{definition}

Keeping in mind that $\Phi_F \circ \Phi_F=F$ (that is $\Phi_F$ is
an involution \cite{Fuster2}), the general conversion method can
be described in the following steps:

\textit{Input: }A nonlinear Boolean function in ANF, e.g.
\[
F(m_{0},\,m_{1},%
\,m_{2})=m_{2}\,m_{0}\oplus m_{2}\,m_{1}\oplus m_{1}.
\]
\begin{itemize}
\item  \textit{Step 1:} Define $\Phi_F$
\[
\Phi_F(m_{0},\,m_{1},%
\,m_{2})=M_{101}\oplus M_{110}\oplus M_{010}\,.
\]

\item  \textit{Step 2:} Substitute every minterm by its
corresponding minterm function and cancel common terms
\[
\begin{array}{lll}
\Phi_F(m_{0},\,m_{1},%
\,m_{2}) & =& (m_{2}\,m_{1}\,m_{0}\oplus m_{2}\,m_{0})\,\oplus  \\
&  & (m_{2}\,m_{1}\,m_{0}\oplus m_{2}\,m_{1})\,\oplus  \\
&  & (m_{2}\,m_{1}\,m_{0}\oplus m_{2}\,m_{1}\oplus
m_{1}\,m_{0}\oplus m_{1}) \\
& = & \,\,\, m_{2}\,m_{1}\,m_{0}\oplus m_{2}\,m_{0}\oplus
m_{1}\,m_{0}\oplus m_{1}\,.
\end{array}
\]

\item  \textit{Step 3:} Apply the function $\Phi_F$ again
\[
\Phi_F \circ \Phi_F(m_{0},\,m_{1},%
\,m_{2})=F(m_0,m_1,m_2)=M_{111}\oplus M_{101}\oplus M_{011}\oplus
M_{010}\,.
\]
\end{itemize}

\textit{Output: }$F$ expressed in terms of its minterms.

The number of minterms in step $3$ gives the number of $1 $'s in
the generated sequence. Notice that such a number equals the
number of non-cancelled terms in step $2$. This will be the
criterium followed in the next section.

\section{A Binary Model to Calculate the Degree of Balancedness in LFSR-Combinational Generators}
A computational procedure that automates the comparison among
different minterm functions in $\Phi_F $, checks the cancelled
terms and computes the number of final terms is presented. Such a
procedure is based on an $L$-bit string representation.

\subsection{Additional Definitions for the Computation}
For the sake of simplicity, the functions $F$ and $\Phi_F$ will be
written as $F=\bigoplus\limits_{i}m_{\alpha _{i}}$ and $\Phi
_{F}=\bigoplus\limits_{i}M_{\alpha _{i}}$, respectively, where
$\bigoplus\limits_{i}$ represents the exclusive-OR sum on the
index $i$.

The own definition of minterm offers us a natural representation
regarding the computational procedure. In fact, every minterm
$M_{\alpha}$ is represented by an $L$-bit string numbered
$0,1,...,L-1$ from right to left. As before, if the \textit{i-th}
variable $m_i$ is in its true form, then the \textit{i-th} bit of
such a string takes the value $1$; otherwise, the value will be
$0$. According to equation (\ref{equation:1}), the number of terms
in the minterm function of $M_{\alpha}$ is $2^{L-d\left( \alpha
\right) }$ terms, $d\left( \alpha \right) $ being the number of $1^{\prime }s$ in $%
\alpha $.

\begin{definition}
We call \textit{maximum common development} (mcd) of two minterms $%
M_{\alpha}$ and $M_{\beta}$, notated $MD\left( M_{\alpha},M_{\beta
}\right) $, to the minterm $M_{\chi}$ such that $\chi =\alpha \cup
\beta $.
\end{definition}
Under this $L$-bit string representation, the mcd can be realized
by means of a bit-wise OR operation between the binary strings of
both minterms. The mcd represents all the terms that the minterm
functions of $M_{\alpha}$ and $M_{\beta}$ have in common. For
instance, for an LFSR of $L=4$ stages and given two minterm
functions:
\[
\begin{array}{lll}
M_{0011} & = & m_{3}\,m_{2}\,m_{1}\,m_{0}\oplus m_{3}\,m_{1}\,m_{0}\oplus m_{2}\,m_{1}\,m_{0}\oplus  m_{1}\;m_{0}\\
M_{1001} & = & m_{3}\,m_{2}\,m_{1}\,m_{0}\oplus
m_{3}\,m_{2}\,m_{0}\oplus m_{3}\;m_{1}\,m_{0}\oplus m_{3}\,m_{0}
\end{array}
\]
the mcd is:
\[
MD\left(M_{0011},M_{1001}\right)
=M_{1011}=m_{3}\,m_{2}\,m_{1}\,m_{0}\oplus m_{3}\,m_{1}\,m_{0}\, ,
\]
that is to say the common terms to both functions.

If the two minterm functions of $M_{\alpha}$ and $M_{\beta}$ are
added, $M_{\alpha}\oplus M_{\beta}$, then the terms corresponding
to their mcd are cancelled. Thus, the total number of terms in
$M_{\alpha}\oplus M_{\beta}$ is the number of terms in
$M_{\alpha}$ plus the number of terms in $M_{\beta}$
minus twice the number of terms in the mcd, that is $%
2^{L-d\left( \alpha \right) }+2^{L-d\left( \beta \right) }-2\cdot
2^{L-d\left( \alpha \cup \beta \right) }$.

Then, we introduce a new kind of auxiliary function notated $H$.
\begin{definition}
The auxiliary function $H$ is defined as $H=\bigoplus\limits_i
s_iM_{\alpha _i}$, $s_i$ being an integer with sign that specifies
how many times $M_{\alpha_i}$ is contained in $H$.
\end{definition}
In a symbolic way, $H$ indicates whether the minterm functions
$M_{\alpha _i}$ are added (sign +) or cancelled (sign -) as well
as how many times such functions have been added or cancelled. A
function $H$ is a special way of recording what happens when
different minterm functions are operated.

Finally, the mcd can be applied to the functions $H$ too.
\begin{definition}
Given two auxiliary functions $H_l=\bigoplus\limits_is_iM_{\alpha
_i}$ and $H_m=\bigoplus\limits_js_jM_{\beta _j}$, the maximum
common development is defined as
\begin{equation}\label{equation:2}
MD\left( H_l,H_m\right)
=\bigoplus\limits_i\bigoplus\limits_j\left( s_is_j\right) MD\left(
M_{\alpha _i},M_{\beta _j}\right)
=\bigoplus\limits_i\bigoplus\limits_j\left( s_is_j\right)
M_{\alpha _i\cup \beta _j}
\end{equation}
where $s_is_j$ is the integer product between both factors.
\end{definition}

As before the mcd of functions $H$ represents the common terms to
both functions. Next, the computational method is described.

\subsection{Computation of the Number of 1's in the Output Sequence for Generators with one LFSR}

Let $F=\bigoplus\limits_{i}m_{\alpha _i}\;(i=1,...,N)$ be a
nonlinear Boolean function of $N$ terms applied to the $L$ stages
of an LFSR. The aim of this section is to compute the number of
$1^{\prime }s$, notated $\mathcal{U}_F$, in the sequence generated
by $F$. Throughout the computation, the most complicated feature
is to know what minterms and how many times are repeated (or
cancelled) without keeping an exhaustive record of them. For
example, let $M_{\alpha_1}$, $M_{\alpha_2}$ and $M_{\alpha_3}$ be
minterms in $\Phi_F$. Indeed, $M_{\alpha _1\cup \alpha _2}$ are
the common terms to $M_{\alpha_1}$ and $M_{\alpha_2}$ that have to
be eliminated. In addition, $M_{\alpha _1\cup \alpha _3}$ have the
same meaning concerning $M_{\alpha_1}$ and $M_{\alpha_3}$. The
question that arises in a natural way is: what happens with the
terms of $M_{\alpha _1\cup \alpha _2}$ that are repeated in
$M_{\alpha _1\cup \alpha _3}$ but that have been previously
cancelled? A similar situation occurs when $M_{\alpha _2\cup
\alpha _3}$ is compared with $M_{\alpha _1\cup \alpha _2}$ and
$M_{\alpha _1\cup \alpha _3}$. The situation becomes worse and
worse when more minterm functions are involved.

In fact, when $M_{\alpha _i}$ an arbitrary minterm of $\Phi_F$ is
considered, then the auxiliary function $H_i$ keeps count of the
minterms involved so far as well as it keeps count of which of
them are repeated or cancelled. At the end of the procedure, when
the last minterm $M_{\alpha _N}$ is introduced, then the auxiliary
function $H_N$ includes all the final minterms with their
corresponding signs. The last step of this computation is just to
determine the number of $1^{\prime }s$ in the output sequence by
counting the number of terms included in each minterm
$M_{\alpha_j}$ of $H_N$ (see subsection 2.4). According to
equation (\ref{equation:1}), this number is given by $2^{L-d\left(
\alpha_j \right) }$. So keeping in mind all these considerations,
the following computational procedure is introduced:
\vspace*{0.2cm}

\textit{Input: }A nonlinear Boolean function in ANF, e.g.
$F=\bigoplus\limits_{i}m_{\alpha _i}\;(i=1,...,N)$.
\begin{itemize}
\item  \textit{Step 1:} Define the function $\Phi_F$ from the $N$
terms $m_{\alpha _i}$ of $F$. Initialize the function $H_0$ with a
null value,
\begin{equation}\label{equation:3}
H_0=\oslash \,.
\end{equation}

\item  \textit{Step 2:} For $i=1,\, \ldots , \,N$
\begin{equation}\label{equation:4}
H_i=H_{i-1}+M_{\alpha _i}-2\cdot MD\left( M_{\alpha
_i},H_{i-1}\right)\, .
\end{equation}

\item  \textit{Step 3:} From the final form of
$H_N=\sum\limits_js_jM_{\beta _j}$, compute the number of
$1^{\prime }s$ in the generated sequence by means of the
expression
\begin{equation}\label{equation:5}
\mathcal{U}_F =\sum\limits_js_j\cdot 2^{L-d\left( \beta _j\right)
}\,.
\end{equation}
\end{itemize}

\textit{Output: }$\mathcal{U}_F$ that is the number of $1^{\prime
}s$ in the output sequence.

 In this way, the number of
$1^{\prime }s$ is obtained as an exponential function in the
length of the LFSR. Comparing $\mathcal{U}_F$ with the expected
value ($\mathcal{U}_F \approx T/2$), the acceptance/rejection
criterium can be applied.

\subsection{Computation of the Number of 1's in the Output Sequence for Generators with more than one LFSR}
For several LFSRs $A,\,B,...,\,Z$ of lengths
$L_{A},$ $\,L_{B},\,...\,,\,L_{Z}$ respectively with $L=L_{A}+\,L_{B}+\,...+%
\,L_{Z}$, the procedure is analogous to the previous one except
for:
\begin{enumerate}
    \item The bits of the $L$-bit strings are assigned to the
    different LFSRs such as follows: the first $L_{A}$ bits (from
    right to left) correspond to the representation of the
    individual minterms of the LFSR $A$. The next $L_{B}$ bits correspond to the representation of the
    individual minterms of the LFSR $B$ ... the last $L_{Z}$ bits correspond to the representation of the
    individual minterms of the LFSR $Z$
    \item Now each $d(\beta_j)$, the number of $1^{\prime }s$ in the
    $j$-th $L$-bit string of $\mathcal{U}_F$ can be expressed
    as:
    \begin{equation}\label{equation:6}
    d(\beta_j)=d_{A}(\beta_j)+d_{B}(\beta_j)+\,...+\,d_{Z}(\beta_j)
    \end{equation}
    where $d_{I}(\beta_j)$ is the number of $1^{\prime }s$ in the
    bits assigned to the $I$-th LFSR. Thus, the number of $1^{\prime }s$
    in the output sequence $\mathcal{U}_F$ can be trivially rewritten
    as:
    \begin{equation}\label{equation:7}
    \mathcal{U}_F =\sum\limits_js_j \cdot ( 2^{L_{A}-d_{A}\left(\beta
    _j\right) } \cdot 2^{L_{B}-d_{B}\left(\beta
    _j\right) }  \, \ldots \,  2^{L_{Z}-d_{Z}\left(\beta
    _j\right) } )
    \end{equation}
\end{enumerate}
Remark that if a minterm of $\Phi_F$ does not contain any
individual minterm, then the corresponding bits in the $L$-bit
string are filled with zeros and the string includes the $2^{L_i}
- 1$ individual minterms of the corresponding LFSR.

\subsection{An Example}
Let $A,B,C$ be three LFSRs of lengths $L_A,L_B,L_C$ respectively.
The LFSR-combinational generator is chosen:
\[
F(a_0,b_0,c_0)=\bigoplus\limits_{i=1}^{3}m_{\alpha
_i}=a_0b_0\oplus b_0c_0\oplus c_0 \, ,
\]
which corresponds to the Geffe's generator \cite{Si}. In order to
fix the $L$-bit strings, very low values are assigned to the
lengths of the LFSRs: $L_A=2,L_B=3,L_C=5$, with ($(L_i,L_j)=1,\;
i\neq j$), thus $L=10$. According to the previous subsection, we
proceed:

\begin{quote}
\textbf{Step 1.-} $\Phi_F(a_0,b_0,c_0) =
\bigoplus\limits_{i=1}^{3}M_{\alpha _i}=A_0\,B_0 \oplus B_0\,C_0
\oplus C_0$ and the minterms of $\Phi_F $ in $10$-bit string
format are:
\[
\begin{array}{lllll}
M_{\alpha _1} & = & A_0\,B_0 & = & 00000\,\,001\,\,01 \\
M_{\alpha _2} & = & B_0\,C_0 & = & 00001\,\,001\,\,00 \\
M_{\alpha _3} & = & C_0 & = & 00001\,\,000\,\,00 \,.
\end{array}
\]
$H$ is initialized $H_0=\oslash \, .$

\textbf{Step 2.-} For $i=1,\, \ldots , \,3$.

$i=1$.- The mcd of $M_{\alpha _1}$ and $H_0$ is
\[
MD\left( M_{\alpha _1},H_0\right) =MD\left( 00000\,\,001\,\,01,
\oslash \right) =\oslash
\]
and the updated $H$ is
\[
H_1=H_0+M_{\alpha _1}-2\cdot MD\left( M_{\alpha _1},H_0\right)
=00000\,\,001\,\,01.
\]
$i=2$.-The mcd of $M_{\alpha _2}$ and $H_1$ is
\[
MD\left( M_{\alpha _2},H_1\right)=00001\,\,001\,\,01.
\]
and the updated $H$ is
\[
H_2=H_1+M_{\alpha _2}-2\cdot MD\left( M_{\alpha _2},H_1\right) =
\]
\[
00000\,\,001\,\,01 \,+\, 00001\,\,001\,\,00  \,-\left[ 2\right]
00001\,\,001\,\,01.
\]
$i=3$.- The mcd of $M_{\alpha _3}$ and $H_2$ is

\[
MD\left( M_{\alpha _3},H_2\right)\,=
\]
\[
00001\,\,001\,\,01 \,+\, 00001\,\,001\,\,00  \,-\left[ 2\right]
00001\,\,001\,\,01 \,=
\]
\[
00001\,\,001\,\,00 \,-\left[ 1\right] 00001\,\,001\,\,01.
\]
and the updated $H_3$ is
\[
H_3 = H_2+M_{\alpha _3}-2\cdot MD\left( M_{\alpha _3},H_2\right) =
\]
\[
00000\,\,001\,\,01 \,+\, 00001\,\,000\,\,00  \,-\left[ 1\right]
00001\,\,001\,\,00.
\]
\textbf{Step 3.-} Calculation of the number of $1^{\prime }s$ from
$H_3$

\[
\begin{array}{lll}
\;\;\;\;\;\;\;\;00000\,\,001\,\,01  & \,\,\,\mbox{implies} &
\;\;2^{L_A-1}\,2^{L_B-1}\,(2^{L_C}-1) \;\,\,\mbox{ones} \\
\;\;\;\;\;\;\;\;00001\,\,000\,\,00   & \,\,\,\mbox{implies} &
\;\;(2^{L_A}-1)\,(2^{L_B}-1)\,2^{L_C-1} \,\,\mbox{ones} \\
-\left[ 1\right] \,00001\,\,001\,\,00  & \,\,\,\mbox{implies}  &
\;\;-\,(2^{L_A}-1)\,2^{L_B-1}\, 2^{L_C-1} \,\,\mbox{ones.}
\end{array}
\]
Thus,
\[
\begin{array}{lll}
\mathcal{U}_F & = &
2^{L_A-1}\,2^{L_B-1}\,(2^{L_C}-1)+(2^{L_A}-1)\,(2^{L_B}-1)\,2^{L_C-1} \\
&  & - \,(2^{L_A}-1)\,2^{L_B-1}\, 2^{L_C-1} \\
& = &
2^{L_A-1}\,2^{L_B-1}\,(2^{L_C}-1)+(2^{L_A}-1)\,(2^{L_B-1}-1)\,2^{L_C-1}.
\end{array}
\]
\end{quote}

For lengths of the LFSRs in, for instance, a cryptographic range
$L_i \approx 64$ the number of $1^{\prime }s$ in the output
sequence is $ \simeq T/2$. Consequently, the generated sequence is
balanced.

The application of this procedure gives us a general expression
for the number of $1^{\prime }s$ in the sequence produced by a
Geffe's generator. From very low values of $L_i$, a general
expression is achieved that can be applied to LFSR lengths in a
range of practical interest. Similar results can be obtained by
applying the previous procedure to other standard
LFSR-combinational generators \cite{Si}.

It must be noticed that $\mathcal{U}_F$ depends exclusively on the
LFSR lengths but not on the connection polynomials. Thus, the same
combinational generator applied to different LFSRs of the same
lengths will give sequences with the same number of binary digits.
Remark that the particular form of these generators, Boolean
functions with a few terms of low orders, allows one the
application of the computational procedure with negligible time
and memory complexity.

\section{Practical Design of Balanced Sequence Generators}
From the previous section, practical rules to design balanced
LFSR-combinational generators can be deduced. The application to
nonlinear filters or nonlinear combination generators is
considered separately.

\subsection{Application of the Computational Model to Nonlinear Filters}
A simple rule to guarantee balancedness in the output sequence of
nonlinear filters is introduced.
\begin{proposition}
Let $F$ be a Boolean function of the form
\[
F(m_{0},\,m_{1},\,...,\,%
\,m_{L-1})=F'(m_{0},\,m_{1},\,...,\,m_{j-1},\,m_{j+1},\,...,%
\,m_{L-1})\oplus m_j,
\]
where $F'$ is an arbitrary Boolean function in $L-1$ variables
given in ANF Then, the output sequence generated by the nonlinear
filter $F$ is balanced.
\end{proposition}

\textbf{Proof:} According to the previous notation
$m_j=m_{\alpha_N}$, then the last term of $\Phi_F $ will be
$M_{\alpha _N}=M_j$. Therefore, at the $N$-th loop, the final
updating of $H$ is:
\[
\begin{array}{lll}
H_N & = & H_{N-1}+M_{\alpha _N}-2\cdot MD\left( M_{\alpha
_N},H_{N-1}\right)
\end{array}
\]
Nevertheless, the terms $H_{N-1}$ and $2\cdot MD\left( M_{\alpha
_N},H_{N-1}\right)$ in the second member contains exactly the same
binary strings except for the $j$-th index. Thus, from the point
of view of the number of $1^{\prime }s$ and for every pair of
quasi-equal strings, we have:
\[
\begin{array}{lll}
2 ^{L-k}-2\cdot 2 ^{L-(k+1)}= & 2 ^{L-k}-2 ^{L-k}= & 0 \;\;\;\;
\forall k.
\end{array}
\]
Therefore, the final number of $1^{\prime }s$ in the sequence
generated is due exclusively to the contribution of the term
$M_{\alpha _N}$ and such a contribution is $2^{L-1}$ ones. Thus,
\[
\mathcal{U}_F = 2^{L-1}.
 \]
Consequently, the output sequence of such generators will always
be balanced.

\hfill $\Box$

This result can be directly applied to TOYOCRYPT-HR1 keystream
generator for the stream cipher TOYOCRYPT-HS1 \cite{IPA}. In fact,
this pseudorandom binary generator includes an LFSR of length
$L=128$ where the last stage, notated $m_{127}$, appears as a
unique term with non-repeated index in the Boolean function $F$.
In this practical example and according to the previous algorithm
the balancedness of the output sequence is guaranteed.

\subsection{Application of the Computational Model to Nonlinear Combination Generators}
The procedure explained in subsection 3.3 lets us determine the
number of $1^{\prime }s$ in the sequence obtained from any
nonlinear combination generator. Let us see its application to the
following examples made up of three LFSRs.

\begin{center}
\begin{tabular}{l}
$F_0=a_0b_0\oplus b_0c_0\oplus a_0c_0\oplus a_0\oplus b_0\oplus c_0$ \\
$F_1=a_0b_0\oplus b_0c_0\oplus a_0\oplus b_0\oplus c_0$ \\
$F_2=a_0b_0\oplus b_0c_0\oplus b_0$ \\
$F_3=a_0b_0\oplus b_0c_0\oplus a_0$ \\
$F_4=a_0b_0\oplus c_0$%
\end{tabular}
\end{center}

In all cases the period of the generated sequence is $T=
(2^{L_{A}}-1)(2^{L_{B}}-1)(2^{L_{C}}-1)$ (supposed $(L_i,L_j)=1,\;
i\neq j$).The general expressions that quantify the number of
$1^{\prime }s$ for the above generators are obtained such as
follows:

\textit{Function} $F_0$: the corresponding $H_6$ is:
\[
\begin{array}{llllllll}
H_6 & = & + & \left[ 1\right]00000\,\,000\,\,01 & + & \left[
1\right]00000\,\,001\,\,00 & + & \left[ 1\right]
00001\,\,000\,\,00 \\

& & - & \left[ 1\right]00000\,\,001\,\,01 & - & \left[
1\right]00001\,\,001\,\,00 & - & \left[ 1\right]
00001\,\,000\,\,01. \end{array}
\]

Thus,
\[
\begin{array}{lll}
\mathcal{U}_F & = &
\;\;\;\;2^{L_A-1}\,(2^{L_B}-1)\,(2^{L_C}-1) + \,(2^{L_A}-1)\,2^{L_B-1}\,(2^{L_C}-1)\\
&  & + \,(2^{L_A}-1)\,(2^{L_B}-1)\,2^{L_C-1} - \,2^{L_A-1}\,2^{L_B-1}\,(2^{L_C}-1)\\
& &  - \,(2^{L_A}-1)\,2^{L_B-1}\,2^{L_C-1}-
\,2^{L_A-1}\,(2^{L_B}-1)\,2^{L_C-1}.
\end{array}
\]
\[
\mathcal{U}_F \simeq T/2 \,+ \,T/4.
\]

\textit{Function} $F_1$: the corresponding $H_5$ is:
\[
\begin{array}{llllllll}
H_5 & = & + & \left[ 1\right]00000\,\,000\,\,01 & + & \left[
1\right]00000\,\,001\,\,00 & + & \left[ 1\right]
00001\,\,000\,\,00 \\

& & - & \left[ 1\right]00000\,\,001\,\,01 & - & \left[
1\right]00001\,\,001\,\,00 & - & \left[ 2\right]
00001\,\,000\,\,01 \\

& & + & \left[ 2\right]00001\,\,001\,\,01.
\end{array}
\]

Thus,
\[
\begin{array}{lll}
\mathcal{U}_F & = &
\;\;\;\;2^{L_A-1}\,(2^{L_B}-1)\,(2^{L_C}-1) + \,(2^{L_A}-1)\,2^{L_B-1}\,(2^{L_C}-1)\\
&  & + \,(2^{L_A}-1)\,(2^{L_B}-1)\,2^{L_C-1} - \,2^{L_A-1}\,2^{L_B-1}\,(2^{L_C}-1)\\
& &  - \,(2^{L_A}-1)\,2^{L_B-1}\,2^{L_C-1}-\,2\,\cdot
2^{L_A-1}\,(2^{L_B}-1)\,2^{L_C-1} \\
 & & +\,2\,\cdot2^{L_A-1}\,2^{L_B-1}\,2^{L_C-1}.
\end{array}
\]
\[
\mathcal{U}_F \simeq T/2 \,+ \,T/4.
\]

\textit{Function} $F_2$: the corresponding $H_3$ is:
\[
\begin{array}{llllllll}
H_3 & = & - & \left[ 1\right]00000\,\,001\,\,01 & - & \left[
1\right]00001\,\,001\,\,00 & + & \left[ 2\right]
00001\,\,001\,\,01 \\

& & + & \left[ 1\right]00000\,\,001\,\,00 . \end{array}
\]

Thus,
\[
\begin{array}{lll}
\mathcal{U}_F & = &
- \,2^{L_A-1}\,2^{L_B-1}\,(2^{L_C}-1) - \,(2^{L_A}-1)\,2^{L_B-1}\,2^{L_C-1}\\
&  & + \,2 \cdot 2^{L_A-1}\,2^{L_B-1}\,2^{L_C-1} +
\,(2^{L_A}-1)\,2^{L_B-1}\,(2^{L_C}-1).
\end{array}
\]
\[
\mathcal{U}_F \simeq T/4.
\]

\textit{Function} $F_3$: the corresponding $H_3$ is:
\[
\begin{array}{llllllll}
H_3 & = & - & \left[ 1\right]00000\,\,001\,\,01 & + & \left[
1\right]00001\,\,001\,\,00 & + & \left[ 1\right]
00000\,\,000\,\,01 .
\end{array}
\]

Thus,
\[
\begin{array}{lll}
\mathcal{U}_F & = &
- \,2^{L_A-1}\,2^{L_B-1}\,(2^{L_C}-1)+(2^{L_A}-1)\,2^{L_B-1}\,2^{L_C-1}\\
&  &  + \,2^{L_A-1}\,(2^{L_B}-1)\,(2^{L_C}-1).
\end{array}
\]
\[
\mathcal{U}_F \simeq T/2.
\]

\textit{Function} $F_4$: the corresponding $H_2$ is:
\[
\begin{array}{llllllll}
H_2 & = & + & \left[ 1\right]00000\,\,001\,\,01 & + & \left[
1\right]00001\,\,000\,\,00 & - & \left[ 2\right]
00001\,\,001\,\,01 .
\end{array}
\]

Thus,
\[
\begin{array}{lll}
\mathcal{U}_F & = &
\;\;\;\;2^{L_A-1}\,2^{L_B-1}\,(2^{L_C}-1)+\,(2^{L_A}-1)\,(2^{L_B}-1)\,2^{L_C-1}\\
&  &  - \,2\cdot 2^{L_A-1}\,2^{L_B-1}\,2^{L_C-1}.
\end{array}
\]
\[
\mathcal{U}_F \simeq T/2.
\]

Table 1. shows the number of $1^{\prime }s$ for a particular choice of $%
L_A,L_B,L_C$. In the rightmost column there appears the order of
magnitude of the number of $1^{\prime }s$ when $L_A,L_B,L_C$ take
values in a cryptographic range. In fact, some kinds of Boolean
functions must be avoided:

\begin{itemize}
\item  The generators $F_0,F_1$ include all the first order
terms $a_0,b_0,c_0$ what means a great number of non-cancelled terms in $%
\Phi_F $, consequently their corresponding number of $1^{\prime
}s$ will be greater than $T/2$.

\item  The generator $F_2$ includes $b_0$ in all the terms what
means a great number of cancelled terms in $\Phi_F $, consequently
its corresponding number of $1^{\prime }s$ will be less than
$T/2$.
\end{itemize}

These forms of Boolean functions will never be balanced and
consequently must be rejected. $F_3$ is
similar to $F_2$ except for fact that the linear term $b_0$ does not appear in all the terms. In this way, $F_3$ as well as $%
F_4$, with non-repeated terms, produce sequences with a balanced number of non-cancelled terms in $%
\Phi_F $ what means a number of $1^{\prime }s$ in the range of $%
\simeq T/2$. Thus, the appearance and/or repetition of the
different variables determine the degree of balancedness in the
output sequence.

In brief, a simple study of any particular Boolean function allows
the designer to know whether the considered sequence has a number
of $1^{\prime }s$ in the desired range of magnitude. Small
variations in the form $F$ improve significantly balancedness in
the generated sequence.

\section{Conclusions}

In the present work, a binary model to compute the number of
$1^{\prime }s$ in the sequence obtained from LFSR-combinational
generators has been developed. The computational procedure enables
one to determine the actual degree of balancedness in the output
sequence without generating the whole sequence. The procedure is
based on the \textit{comparison} and \textit{cancellation} of the
common terms in the function $\Phi_F $ of the generator. According
to an adequate binary representation of minterm functions, the
procedure can be carried out by realizing elementary logic
operations on binary strings. The application of the method to
examples with LFSRs of reduced length gives general expressions
for the number of $1^{\prime }s$ in the sequences obtained by
standard generators published in the open literature. Some simple
rules for the design of balanced generators (either nonlinear
filters or combination generators) are also derived.

\vspace*{0.7cm} \noindent \large{\textbf{Acknowledgements}}\\
\normalsize{Work supported by Ministerio de Educaci\'{o}n y
Ciencia (Spain) Projects SEG2004-02418 and SEG2004-04352-C04-03.}

\newpage
\setlength{\tabcolsep}{4pt}
\begin{table}
\begin{center}
\caption{Numerical results for different combination generators}
\label{table:headings1}
\begin{tabular}{cccccrc}
\hline\noalign{\smallskip} & $\;\;L_A$ & $\;\;L_B$ & $\;\;L_C$ &
\,$\;\;Expected\;No.1^{\prime }s$ &
\,$\;\;Actual\;No.1^{\prime }s $ & \,$\;\;Order\;of$\ $magnitude$\\
\noalign{\smallskip} \hline \noalign{\smallskip}
$F_0$ & $7$ & $8$ & $9$ & $8274368$ & $12411328$ & $\simeq T/2+T/4$ \\
\hline $F_1$ & $7$ & $8$ & $9$ & $8274368$ & $12427712$ & $\simeq
T/2+T/4$
\\ \hline $F_2$ & $7$ & $8$ & $9$ & $8274368$ & $4153472$ & $\simeq T/4$
\\ \hline
$F_3$ & $7$ & $8$ & $9$ & $8274368$ & $8314944$ & $\simeq T/2$ \\
\hline
$F_4$ & $7$ & $8$ & $9$ & $8274368$ & $8282368$ & $\simeq T/2$ \\
\hline
\end{tabular}
\end{center}
\end{table}
\setlength{\tabcolsep}{1.4pt}

\end{document}